\documentstyle[11pt]{article}

\def\dalemb#1#2{{\vbox{\hrule height .#2pt
        \hbox{\vrule width.#2pt height#1pt \kern#1pt
                \vrule width.#2pt}
        \hrule height.#2pt}}}

\let\a=\alpha \let\b=\beta   \let\e=\epsilon
  \let\q=\theta  
  \let\n=\nu

      \let\G=\Gamma

\def\nn{\nonumber} \def\bd{\begin{document}} \def\ed{\end{document}}
\def\ds{\documentstyle} \let\fr=\frac \let\bl=\bigl \let\br=\bigr
\let\Br=\Bigr \let\Bl=\Bigl
\let\bm=\bibitem
\let\na=\nabla
\let\pa=\partial \let\ov=\overline
\def\ie{{\it i.e.\ }}
\newcommand{\be}{\begin{equation}}
\newcommand{\ee}{\end{equation}}
\def\ba{\begin{array}}
\def\ea{\end{array}}
\def\ft#1#2{{\textstyle{{\scriptstyle #1}\over {\scriptstyle #2}}}}
\def\fft#1#2{{#1 \over #2}}
\def\del{\partial}
\def\sst#1{{\scriptscriptstyle #1}}
\def\oneone{\rlap 1\mkern4mu{\rm l}}
\def\e7{E_{7(+7)}}
\def\td{\tilde}
\def\wtd{\widetilde}
\def\im{{\rm i}}
\def\bog{Bogomol'nyi\ }
\def\q{{\tilde q}}
\def\hast{{\hat\ast}}
\def\0{{\sst{(0)}}}
\def\1{{\sst{(1)}}}
\def\2{{\sst{(2)}}}
\def\3{{\sst{(3)}}}
\def\4{{\sst{(4)}}}
\def\5{{\sst{(5)}}}
\def\6{{\sst{(6)}}}
\def\7{{\sst{(7)}}}
\def\8{{\sst{(8)}}}
\def\n{{\sst{(n)}}}

\newcommand{\ho}[1]{$\, ^{#1}$}
\newcommand{\hoch}[1]{$\, ^{#1}$}
\newcommand{\bea}{\begin{eqnarray}}
\newcommand{\eea}{\end{eqnarray}}
\newcommand{\ra}{\rightarrow}
\newcommand{\lra}{\longrightarrow}
\newcommand{\Lra}{\Leftrightarrow}
\newcommand{\ap}{\alpha^\prime}
\newcommand{\bp}{\tilde \beta^\prime}
\newcommand{\tr}{{\rm tr} }
\newcommand{\Tr}{{\rm Tr} }
\newcommand{\NP}{Nucl. Phys. }
\newcommand{\tamphys}{\it Center for Theoretical Physics,
Texas A\&M University, College Station, Texas 77843}
\newcommand{\ens}{\it Laboratoire de Physique Th\'eorique de l'\'Ecole
Normale Sup\'erieure\hoch{3}\\
24 Rue Lhomond - 75231 Paris CEDEX 05}

\newcommand{\auth}{H. L\"u\hoch{\dagger},
C.N. Pope\hoch{\ddagger1} and J. Rahmfeld\hoch{*2}}

\begin{document}
\begin{flushright}
\hfill{CTP TAMU-22/98}\\
\hfill{LPTENS-98/22}\\
\hfill{SU-ITP-98/31}\\
\hfill{hep-th/9805151}\\
\hfill{May 1998}\\
\end{flushright}

\vspace{20pt}

\begin{center}
{\large {\bf A Construction of Killing Spinors on $S^n$}}

\vspace{30pt}

\auth

\vspace{15pt}
{\hoch{\dagger}\ens}

\vspace{10pt}
{\hoch{\ddagger}\tamphys}

\vspace{10pt}
{\hoch{*}{\it Department of Physics, Stanford University
Stanford, CA 94305-4060}}

\vspace{40pt}

\underline{ABSTRACT}
\end{center}

    We derive simple general expressions for the explicit Killing
spinors on the $n$-sphere, for arbitrary $n$. Using these results we
also construct the Killing spinors on various AdS$\times$Sphere
supergravity backgrounds, including AdS$_5\times S^5$, AdS$_4\times
S^7$ and AdS$_7\times S^4$.  In addition, we extend previous results
to obtain the Killing spinors on the hyperbolic spaces $H^n$.

{\vfill\leftline{}\vfill
\vskip  10pt
\footnoterule
{\footnotesize
        \hoch{1}        Research supported in part by DOE
grant DE-FG03-95ER40917. \vskip -12pt}  \vskip  14pt
{\footnotesize
        \hoch{2}  Research supported by NSF Grant PHY-9219345.
               \vskip   -12pt}  \vskip  14pt
{\footnotesize
        \hoch{3} Unit\'e Propre du Centre National de la Recherche
Scientifique, associ\'ee \`a l'\'Ecole Normale Sup\'erieure
\vskip -12pt} \vskip 10pt
{\footnotesize \hoch{\phantom{3}} et \`a l'Universit\'e de Paris-Sud.
\vskip -12pt} \vskip 10pt}

\pagebreak
\setcounter{page}{1}

\section{Introduction}

     Finding the explicit form of Killing spinors on curved spaces can
be an involved task. Often, one merely uses integrability conditions
to establish their existence and to determine their multiplicities.
In this way it is easy to show that spheres and anti-de Sitter
spacetimes preserve all supersymmetries, {\it i.e.}\ they admit the
maximum number of Killing spinors.  However, one does not obtain
explicit solutions by this method.  Although establishing their
existence is often sufficient, there are situations where it is
necessary to know their explicit forms.

     There exists a very simple explicit construction of the Killing
spinors on $n$-dimensional anti-de Sitter spacetime AdS$_n$, for
arbitrary $n$ \cite{lpt}.  This exploits the fact that AdS$_n$ can be
written in horospherical coordinates, in terms of which the metric
takes the simple form
\be
ds^2 = dr^2 + e^{2 r}\, \eta_{\a\beta}\, dx^\a\, dx^\beta
\ ,\label{adsmetric}
\ee
where $\eta_{\a\beta}$ is the Minkowski metric in $(n-1)$ dimensions,
and the Ricci tensor satisfies $R_{\mu\nu} =- (n-1)\, g_{\mu\nu}$.  It
was shown in \cite{lpt} that the Killing spinors, satisfying $D_\mu\,
\epsilon = \ft12 \Gamma_\mu\, \epsilon$, are then expressible as
\be
\epsilon = e^{\ft12\, r}\, \epsilon_0^+\ ,\qquad {\rm or}\qquad
\epsilon = \Big( e^{-\ft12\,  r} +
e^{\ft12\, r}\, x^\a\, \Gamma_\a
\Big)\, \epsilon_0^-\ ,
\ee
where $\epsilon_\pm$ are arbitrary constant spinors satisfying
$\Gamma_r\, \epsilon_0^\pm = \pm\, \epsilon_0^\pm$.  One can
alternatively write the two kinds of Killing spinor together in one
equation, as
\be
\epsilon= e^{\ft12\, r\, \Gamma_r}\, \Big( 1 + \ft12\, x^\a\,
\Gamma_\a (1-\Gamma_r)\Big)\, \epsilon_0\ ,\label{adsspinors}
\ee
where $\epsilon_0$ is an arbitrary constant spinor.  It is therefore
manifest that the number of independent Killing spinors is equal to
the number of components in the spinors.  (The Killing spinors for
AdS$_4$, written in the standard AdS coordinate system, were obtained
in \cite{bf}.) It is worth remarking that the horospherical metric
(\ref{adsmetric}) can equally well have other spacetime signatures
$(p,n-p)$, by taking other signatures $(p,n-p-1)$ for the metric
$\eta_{\a\b}$.  The isometry group is $SO(p+1,n-p)$.  The case $p=1$
gives AdS$_n$, with $SO(2,n-1)$, while $p=0$ gives the
positive-definite hyperbolic metric on $H^n$, with $SO(1,n)$.
(Expressions for the Killing spinors on $H^2$ and $H^3$, which are
special cases of (\ref{adsspinors}), were given in \cite{fy}.)  Thus
equation (\ref{adsspinors}) gives the Killing spinors on all of the
AdS$_n$ spacetimes, hyperbolic spaces $H^n$, and the other
maximally-symmetric spacetimes with $(p,n-p)$ signature.

    There is an alternative Killing spinor equation that one can
consider when $n$ is even, namely $D_\mu\,\epsilon = \ft{\im}{2}\,
\gamma\, \Gamma_\mu\, \epsilon$, where $\gamma$ is the chirality
operator, expressed as an appropriate product over the $\Gamma_\mu$,
with $\gamma^2=1$.  We easily see that the solutions of this equation
can be written as
\be
\epsilon= e^{\ft{\im}{2}\, r\, \gamma\, \Gamma_r}\, \Big(
1 + \ft{\im}{2}\, \gamma\, x^\a\, \Gamma_\a (1-\im\,
\gamma\, \Gamma_r)\Big)\, \epsilon_0\ .
\ee

    Note that in all the cases above, we considered a ``unit radius''
AdS$_n$, or $H^n$, {\it etc}., given by (\ref{adsmetric}).  It is
trivial to extend the results to an arbitrary scale size, by replacing
(\ref{adsmetric}) by $ds^2=\lambda^{-2}\, (dr^2 + e^{2r}\,
\eta_{\a\b}\, dx^\a\, dx^\b)$, which has the Ricci tensor $R_{\mu\nu}
= -(n-1)\, \lambda^2\, g_{\mu\nu}$.  The Killing spinor equations then
become $D_\mu\, \epsilon= \ft12\lambda\, \Gamma_\mu\, \epsilon$, {\it
etc}.  It is easily seen that the solutions are given by precisely the
same expressions (\ref{adsspinors}), {\it etc}., with no modifications
whatsoever.  (In \cite{lpt} a different coordinatisation of the
general-radius AdS$_n$ was used, in which the expressions for the
Killing spinors {\it do} depend upon the scale-setting parameter.)

     In this paper, we find an explicit construction of the Killing
spinors on $S^n$.  (Explicit results for $n=2$ and $n=3$ were obtained
in \cite{fy}.) One might think that since AdS$_n$ can be related to
$S^n$ by appropriate complexifications of coordinates, it should be
possible to obtain expressions for the Killing spinors on $S^n$ that
are analogous to those given above.  However, things are not quite so
simple, because the ability to write the metric on AdS$_n$ in the
simple form (\ref{adsmetric}) depends rather crucially on the fact
that its isometry group $SO(2,n-1)$ is non-compact.  (One can easily
see that (\ref{adsmetric}) has $(n-1)$ commuting Killing vectors
$\del_\mu$, which exceeds the rank ${[}(n+1)/2{]}$ of the isometry
group when $n>3$.  This is not possible for compact groups.)  We shall
thus present a different construction for the Killing spinors of
$S^n$, which, although more complicated, is still explicit, and of an
essentially simple structure.  Our main result is contained in
equation (\ref{sol1}) in section 2, which also contains a detailed
proof.  In section 3 we combine the results for AdS and spheres, to
give the explicit expressions for Killing spinors in various
AdS$_m\times S^n$ supergravity backgrounds, with $(m,n)=$ (4,7),
(7,4), (5,5), (3,3), (3,2), (2,3), (2,2).  In appendix A, we collect
some useful expressions for the representation and decomposition of
Dirac matrices.

\section{Killing spinors on $S^n$}

\subsection{Results}

   We begin by writing the metric on a unit $S^n$ in terms of that for
a unit $S^{n-1}$ as
\be
ds_n^2 = d\theta_n^2 + \sin^2\theta_n \,
ds_{n-1}^2\ ,\label{snmetric}
\ee
with $ds_1^2 = d\theta_1^2$.  This has Ricci tensor given by $R_{ij}
= (n-1)\, g_{ij}$.  We then consider the Killing spinor
equation on the unit $n$-sphere, for arbitrary $n$, namely
\be
D_j\, \epsilon = \fft{\im}2\, \Gamma_j\, \epsilon\ .\label{ks1}
\ee
We shall first present our results for the solutions to this equation,
and then present the proof later.  We find that the Killing spinors
can be written as
\be
\epsilon= e^{\ft{\im}2\theta_n\,\Gamma_n}\, \Big(\prod_{j=1}^{n-1}
e^{-\ft12
\theta_j\, \Gamma_{j,j+1}} \Big)\, \epsilon_0\ ,\label{sol1}
\ee
where $\epsilon_0$ is an arbitrary constant spinor, and the indices on
the Dirac matrices are vielbein indices.  We use the convention that
the $\Gamma$ matrices are Hermitean, satisfying the Clifford algebra
$\{\Gamma_i,\Gamma_j\} = 2\delta_{ij}$.  Note that here, and in all
other analogous formulae in the paper, the factors in the product in
(\ref{sol1}) are ordered anti-lexigraphically, {\it i.e.}\ starting
with the $\theta_{n-1}$ term at the left.  Note also that the
exponential factors in (\ref{sol1}) can be written as
\be
e^{\ft{\im}2\theta_n \,\Gamma_n} = \oneone\, \cos\ft12\theta_n
   + i\Gamma_n\, \sin\ft12\theta_n\ ,\qquad
e^{-\ft12 \theta_j\, \Gamma_{j,j+1}} = \oneone\, \cos\ft12\theta_j
-\Gamma_{j,j+1}\, \sin\ft12\theta_j\ .
\ee

     One can also consider the Killing spinor equation with the
opposite sign for the $\Gamma_j$ term, namely
\be
D_j\, \epsilon_- = -\fft{\im}2\, \Gamma_j\, \epsilon_-\ .\label{ks1m}
\ee
The previous solution (\ref{sol1}) is easily modified to give
solutions of this equation. One finds
\be
\epsilon_-= e^{-\ft{\im}2\theta_n\,\Gamma_n}\,
\Big(\prod_{j=1}^{n-1} e^{-\ft12
\theta_j\, \Gamma_{j,j+1}} \Big)\, \epsilon_0\ .\label{sol1m}
\ee
This is immediately verified by noting that (\ref{ks1m}) is obtained
from (\ref{ks1}) by changing the sign of the gamma matrices.

    The Killing spinors discussed above exist on $S^n$ for any $n$.
When $n$ is even, there is an alternative equation that can also be
considered, namely
\be
D_j\, \epsilon = \ft12\, \gamma\, \Gamma_j\, \epsilon\ ,\label{ks2}
\ee
where $\gamma$ is the chirality operator formed from the product of
the $\Gamma$ matrices, satisfying $\gamma^2=1$.  In this case, we find
that the corresponding Killing spinors can be written as
\be
\epsilon= e^{\ft{1}2\theta_n\,\gamma\, \Gamma_n}\,
\Big( \prod_{j=1}^{n-1} e^{-\ft12
\theta_j\, \Gamma_{j,j+1}}\Big)\, \epsilon_0\ ,\label{sol2}
\ee
We may again also consider the Killing spinors satisfying (\ref{ks2})
with the sign of the $\Gamma_j$ term reversed,
namely
\be
D_j\, \epsilon = -\ft12\, \gamma\, \Gamma_j\, \epsilon\ .\label{ks2m}
\ee
The solutions are again obtained by sending $\theta_n\rightarrow
-\theta_n$, giving
\be
\epsilon_- = e^{-\ft{1}2\theta_n\,\gamma\, \Gamma_n}\,
\Big( \prod_{j=1}^{n-1} e^{-\ft12
\theta_j\, \Gamma_{j,j+1}}\Big)\, \epsilon_0\ ,\label{sol2m}
\ee

    As in the AdS and $H^m$ cases discussed in section 1, we may again
trivially extend the results to an $n$-sphere of arbitrary radius,
with metric $ds_n^2 = \lambda^{-2}\, (d\theta_n^2 + \sin^2\theta_n\,
ds_{n-1}^2)$ and Ricci tensor $R_{ij} = (n-1)\, \lambda^2\, g_{ij}$.
The Killing spinor equations are modified to $D_j\,\epsilon =
\ft{\im}{2}\, \lambda\, \Gamma_j\, \epsilon$, {\it etc}., but again
the expressions (\ref{sol1}), {\it etc}. for the Killing spinors
receive no modification whatsoever.

\subsection{Proofs}

     The proofs of these results proceed by substituting our
expressions into the corresponding Killing spinor equations.  We begin
by showing that in the orthonormal basis $e^n=d\theta_n$, $e^a=
\sin\theta_n\, e_{(n-1)}^a$, the spin connection for the metric
(\ref{snmetric}) is given by
\be
\omega^{ab} = \omega^{ab}_{(n-1)}\ ,\qquad \omega^{a,n} =
\cos\theta_n\, e_{(n-1)}^a\ ,
\ee
where $a\le n-1$, and $e_{(n-1)}^a$ and $\omega^{ab}_{(n-1)}$ are the
vielbein and spin connection for $S^{n-1}$.  (Note that the index $n$
always denotes the specific value $n$ of the dimension of the
$n$-sphere.)  Thus we can write the vielbein and spin connection on
$S^n$ as
\bea
e^j &=& \Big(\prod_{k=j+1}^n \sin\theta_k\Big)\, d\theta_j\ ,\nn\\
\omega^{jk} &=& \cos\theta_k\,\Big(\prod_{\ell=j+1}^{k-1}
\sin\theta_\ell\Big) \, d\theta_j\ ,\qquad 1\le j < k\le n\
.\label{eomega}
\eea
The Killing spinor equation (\ref{ks1}) can be written as
\be
\del_j\, \epsilon +\ft14 \omega_j{}^{k\ell}\,
\Gamma_{k\ell}\, \epsilon = \fft{\im}{2}\, e_j{}^k\, \Gamma_k\,
\epsilon\ ,\label{kscomp1}
\ee
where $\omega_j{}^{k\ell}$ and $e_j{}^k$ are the coordinate-index
components of $\omega^{k\ell}$ and $e^k$, {\it i.e.}\ $\omega^{k\ell}=
\omega_j{}^{k\ell}\, d\theta_j$ and $e^k=e_j{}^k\, d\theta^j$.  These
can be read off from (\ref{eomega}).  Note that the indices on the
$\Gamma$ matrices in (\ref{kscomp1}) are vielbein indices.

     We now make the following two definitions:
\bea
U_j{}^k &\equiv & \Big(\prod_{\ell=j+1}^k e^{-\ft12 \theta_\ell\,
\Gamma_{\ell,\ell+1}}\Big)\, \Gamma_{j,j+1}\,
\Big(\prod_{\ell=j+1}^k e^{-\ft12 \theta_\ell\,
\Gamma_{\ell,\ell+1}}\Big)^{-1}\ ,\qquad k\ge j\label{id1}\\
V_j &\equiv & e^{\ft{\im}{2}\theta_n\, \Gamma_n}\, U_j{}^{n-1}\,
e^{-\ft{\im}{2}\theta_n\, \Gamma_n}\ ,\label{id2}
\eea
where as usual, the factors with the larger $\ell$ values in the
product sit to the left of those with smaller $\ell$ values.  (Note
that if the upper limit on the product is less than the lower limit,
then it is defined to be 1.)  It is now evident that verifying that
the expression (\ref{sol1}) gives a solution to the Killing spinor
equation (\ref{ks1}) amounts to proving that
\be
V_j = -\im\, e_j{}^j\, \Gamma_j + \sum_{k>j}^n\, \omega_j{}^{jk}\,
\Gamma_{jk} \ .\label{rtp}
\ee
We prove this by first establishing two lemmata.  The first, whose
proof is elementary, states
that if $X$ and $Y$ are matrices such that ${[} X, Y{]}= 2Z$, and
${[} X,Z{]} = -2Y$, then
\be
e^{\ft12\theta\, X}\, Y\, e^{-\ft12\theta\, X} = \cos\theta\, Y +
\sin\theta\, Z\ .\label{lemma1}
\ee
The second lemma states that
\be
U_j{}^k = \sec\theta_{k+1}\, \omega_j{}^{j,k+1}\, \Gamma_{j,k+1}
 + \sum_{\ell>j}^k \omega_j{}^{j\ell}\, \Gamma_{j\ell}\ , \qquad
k\ge j \ .\label{lemma2}
\ee
We prove this by induction.  From the definition (\ref{id1}), we know
that $U_j{}^j = \Gamma_{j,j+1}$, which clearly satisfies (\ref{lemma2})
since $\omega_j{}^{j,j+1} = \cos\theta_{j+1}$.  Assuming then that
(\ref{lemma2}) holds for a specific $k\ge j$, we will have that
\bea
U_j{}^{k+1} &\equiv& e^{-\ft12 \theta_{k+1}\, \Gamma_{k+1,k+2}}\,
 U_j{}^k\, e^{+\ft12 \theta_{k+1}\, \Gamma_{k+1,k+2}}\ ,\nn\\
&=& \sec\theta_{k+1}\, \omega_j{}^{j,k+1}\,
e^{-\ft12 \theta_{k+1}\, \Gamma_{k+1,k+2}}\, \Gamma_{j,k+1}\,
e^{\ft12 \theta_{k+1}\, \Gamma_{k+1,k+2}} +
\sum_{\ell>j}^k\, \omega_j{}^{j\ell}\, \Gamma_{j\ell}\ ,
\eea
where we have made use of the fact that the $\Gamma_{j\ell}$ in the
last term all commute with $\Gamma_{k+1,k+2}$, since $\ell\le k$.  The
first term can be evaluated using lemma 1, giving
\bea
U_j{}^{k+1} &=& \sec\theta_{k+1}\, \omega_j{}^{j,k+1}\, \Big(
\cos\theta_{k+1} \, \Gamma_{j,k+1} + \sin\theta_{k+1}\, \Gamma_{j,k+2}
\Big) + \sum_{\ell> j}^k \omega_j{}^{j\ell}\, \Gamma_{j\ell}\ ,\nn\\
&=& \tan\theta_{k+1}\, \omega_j{}^{j,k+1}\, \Gamma_{j,k+2} +
\sum_{\ell> j}^{k+1} \omega_j{}^{j\ell}\, \Gamma_{j\ell}\ .
\eea
Now, it follows from (\ref{eomega}) that $\omega_j{}^{j,k+2} =
\cos\theta_{k+2}\, \tan\theta_{k+1}\, \omega_j{}^{j,k+1}$.  Using
this, we then obtain (\ref{lemma2}) with $k$ replaced by $k+1$,
completing the inductive proof.

    Having established the lemmata, we can substitute the expression
$U_j{}^{n-1}$ from (\ref{lemma2}) into the definition of $V_j$ given
in (\ref{id1}), giving
\bea
V_j&=& \sec\theta_n\, \omega_j{}^{jn}\, e^{\ft{\im}{2}\theta_n\,
\Gamma_n}
\, \Gamma_{jn}\, e^{-\ft{\im}{2}\theta_n\,  \Gamma_n} +
\sum_{\ell>j}^{n-1}
\omega_j{}^{j\ell}\, \Gamma_{j\ell}\ ,\nn\\
&=& -\im\, \tan\theta_n\, \omega_j{}^{jn}\, \Gamma_j +
 \sum_{\ell>j}^n \omega_j{}^{j\ell}\,\Gamma_{j\ell}\ ,\label{ppp}
\eea
where we have used lemma 1 to derive the second line.  Since $e_j{}^j
= \tan\theta_n\, \omega_j{}^{jn}$, as can be seen from (\ref{eomega}),
it follows that (\ref{ppp}) gives (\ref{rtp}). This completes the
proof that (\ref{sol1}) satisfies the Killing spinor equation
(\ref{ks1}).  An essentially identical proof shows that (\ref{sol2})
satisfies the alternative Killing spinor equation (\ref{ks2}) in even
dimensions.

\section{Killing spinors on AdS$\times$Sphere}

   An application of the formulae obtained in this paper is to
construct the explicit forms of the Killing spinors in the full
$D$-dimensional spacetime of a supergravity theory that admits an
AdS$_m\times S^n$ solution, where $D=m+n$.  Consider, for example, the
AdS$_4\times S^7$ solution of $D=11$ supergravity.  This is obtained
by taking $F_{\mu\nu\rho\sigma}=6m\, \epsilon_{\mu\nu\rho\sigma}$ with
$\mu=0,8,9,10$, implying that the Ricci tensors on AdS$_4$ and $S^7$
satisfy $R_{\mu\nu}= -12m^2\, g_{\mu\nu}$ and $R_{mn}=6m^2\, g_{mn}$
respectively \cite{dnp}.  The Killing spinors $\epsilon$ must satisfy
\be
0=\delta\psi_{\sst M} = D_{\sst M}\, \epsilon -
\ft1{288}(\hat\Gamma_{\sst{MNPQR}}\, F^{\sst{NPQR}}
-8 F_{\sst{MNPQ}}\, \hat\Gamma^{\sst{NPQ}})\, \epsilon\ .
\ee
Using the appropriate decomposition of Dirac matrices given in
appendix A, this implies that on AdS$_4$ and $S^7$ we must have
\bea
\hbox{AdS}_4:&& D_\mu\, \epsilon_{\sst{\rm AdS}}= \im\, m\,
\gamma\, \Gamma_\mu\,
\epsilon_{\sst{\rm AdS}}\ ,\nn\\
S^7:&& D_j\, \eta = \ft{\im}2\, m\, \Gamma_j\, \eta\ \label{AdS4S7}
\eea
with $j=1,\ldots ,7$.  From the results obtained in this paper we
find that the Killing spinors on AdS$_4\times S^7$ can be written as
\be
\hbox{AdS}_4\times S^7:\quad
\epsilon=e^{\ft{\im}{2}\, r\, \hat\gamma\, \hat \Gamma_r}\, \Big( 1 +
\ft12 x^\a(\im\,
\hat\gamma\, \hat\Gamma_\a + \hat\Gamma_r\,\hat\Gamma_\a)\Big)\,
e^{\ft{\im}2\, \theta_7\, \hat\gamma\, \hat\Gamma_7}\,
\Big(\prod_{j=1}^6 e^{-\ft12\theta_j\, \hat\Gamma_{j,j+1}} \Big)\,
\epsilon_0\ ,
\ee
where $\hat\gamma \equiv -\ft{\im}{24}\, \epsilon^{\mu\nu\rho\sigma}\,
\hat\Gamma_{\mu\nu\rho\sigma}= \gamma\otimes \oneone$ is a ``pseudo
chirality operator,'' and $\epsilon_0$ is an arbitrary 32-component
constant spinor in $D=11$.  Note that the explicit
numerically-assigned indices refer to the seven directions on the
7-sphere.\footnote{Note that as implied by (\ref{AdS4S7}), the AdS$_4$
and $S^7$ have different radii, which are related by the
eleven-dimensional field equations.  However, as noted before, in our
coordinatisation the Killing spinors are independent of the scale
sizes.}

  In $D=11$ supergravity there is also a solution AdS$_7\times S^4$.
An analogous calculation gives the result that the Killing spinors in
this background can be written as
\be
\hbox{AdS}_7\times S^4:\quad
\epsilon=e^{\ft{1}{2}\, r\, \hat\gamma\, \hat \Gamma_r}\, \Big( 1 +
\ft12 x^\a(
\hat\gamma\, \hat\Gamma_\a + \hat\Gamma_r\,\hat\Gamma_\a)\Big)\,
e^{\ft{1}2\, \theta_4\, \hat\gamma\, \hat\Gamma_4}\,
\Big(\prod_{j=1}^3 e^{-\ft12\theta_j\, \hat\Gamma_{j,j+1}} \Big)\,
\epsilon_0\ ,
\ee
where $\hat \gamma \equiv \hat \Gamma_{1234}=\oneone\otimes \gamma$,
and all numerically-assigned
indices refer to the four directions on $S^4$.  Again, $\epsilon_0$ is
an arbitrary 32-component constant spinor in $D=11$.

  As another explicit example let us
look at Type IIB supergravity on AdS$_5 \times S^5$. The gravitino
transformation rules are
\be
0=\delta\psi_{\sst M} = D_{\sst M}\, \epsilon
+\ft\im{1920}\, \hat\Gamma^{\sst{NPQRS}}\,\hat\Gamma_{\sst M}
                F_{\sst{NPQRS}}\, \epsilon\ , \label{2b}
\ee
where $\epsilon $ is a ten dimensional spinor of positive chirality,
satisfying
\be
\hat\G_0...\hat\G_9 \epsilon=\epsilon. \label{chirality}
\ee
Choosing now $F_{\mu\nu\rho\lambda\sigma}=4 m\,
\epsilon_{\mu\nu\rho\lambda
\sigma}$ and $F_{ijk\ell m} = 4m\, \epsilon_{ijk\ell m}$,
equation (\ref{2b}) reduces to
\be
D_{\sst M}\, \epsilon -\ft{1}{2}\,m\,  \left(\sigma_1\times
\oneone \times \oneone
\right)\,
\hat\Gamma_{\sst{M}} \, \epsilon\ =0\ ,
\ee
where we are using the (odd,odd) decomposition of Dirac matrices
given in appendix A. With the ansatz
\be
\epsilon=\pmatrix {1 \cr 0}\otimes \epsilon_{\sst{\rm AdS}}\
\otimes \eta
\ee
for a spinor of positive chirality,
we obtain the equations for the AdS$_5$ and $S^5$
subspaces:
\bea
\hbox{AdS}_5:&& D_\mu\, \epsilon_{\sst{\rm AdS}}=  \frac{1}{2}m\,
\Gamma_\mu\, \epsilon_{\sst{\rm AdS}}\ ,\nn\\
S^5:&& D_j\, \eta = \frac{\im}{2}\ m\, \Gamma_j\, \eta\ ,
\eea
which are the standard Killing spinor equations.
Putting the AdS and $S^n$ results together,
we obtain the explicit expression for the Killing spinors on
AdS$_5\times S^5$
\be
\hbox{AdS}_5\times S^5:\quad
\epsilon=e^{\ft{\im}{2} r\, \hat\gamma\, \hat \Gamma_r}\,
\left( 1 +\ft{1}{2}
              x^\a \left(
\im\, \hat\gamma\hat\Gamma_\a + \hat\Gamma_r\, \hat \Gamma_\a
\right)\right)
e^{-\ft{\im}2\, \theta_5\, \hat\gamma\, \hat\Gamma_5}\,
\Big(\prod_{j=1}^4 e^{-\ft12\theta_j\, \hat\Gamma_{j,j+1}} \Big)\,
\epsilon_0  \ ,
\ee
where $\epsilon_0$ is an arbitrary 32-component constant
spinor of positive chirality, and $\hat\gamma\equiv
\hat\Gamma^{12345}=-\sigma_2\otimes\oneone\otimes\oneone$, where the
numerical indices lie in $S^5$.

Four further analogous examples that arise in lower-dimensional
supergravities are
\bea
\hbox{AdS}_3\times S^3:&&
\epsilon=e^{-\ft{\im}{2} r\, \hat{\td\gamma} \hat \Gamma_r}\,
\left( 1 +\ft{1}{2}
              x^\a \left(-\im\, \hat{\td\gamma}\,\hat\Gamma_\a
+ \hat\Gamma_r\, \hat\Gamma_\a \right)\right)
e^{-\ft{\im}2\, \theta_3\,\hat\gamma\,  \hat\Gamma_3}\,
\Big(\prod_{j=1}^2 e^{-\ft12\theta_j\, \hat\Gamma_{j,j+1}} \Big)\,
\epsilon_0  \ ,\nn\\
\hbox{AdS}_3\times S^2:&&
\epsilon=e^{\ft{1}{2}\, r\, \hat\gamma\, \hat \Gamma_r}\, \Big( 1 +
\ft12 x^\a(
\hat\gamma\, \hat\Gamma_\a + \hat\Gamma_r\,\hat\Gamma_\a)\Big)\,
e^{\ft{1}2\, \theta_2\, \hat\gamma\, \hat\Gamma_2}\,
e^{-\ft12\theta_1\, \hat\Gamma_{12}}\,
\epsilon_0\ ,\nn\\
\hbox{AdS}_2\times S^3: &&
\epsilon=e^{\ft{\im}{2}\, r\, \hat\gamma\, \hat \Gamma_r}\, \Big( 1 +
\ft12 x(\im\,
\hat\gamma\, \hat\Gamma_x + \hat\Gamma_r\,\hat\Gamma_x)\Big)\,
e^{\ft{\im}2\, \theta_3\, \hat\gamma\, \hat\Gamma_3}\,
\Big(\prod_{j=1}^2 e^{-\ft12\theta_j\, \hat\Gamma_{j,j+1}} \Big)\,
\epsilon_0\ ,\nn\\
\hbox{AdS}_2\times S^2: &&
\epsilon=e^{\ft{\im}{2}\, r\, \hat\gamma\, \hat \Gamma_r}\, \Big( 1 +
\ft12 x(\im\,
\hat\gamma\, \hat\Gamma_x + \hat\Gamma_r\,\hat\Gamma_x)\Big)\,
e^{\ft{\im}2\, \theta_2\, \hat\gamma\, \hat\Gamma_2}\,
e^{-\ft12\theta_1\, \hat\Gamma_{12}}\,
\epsilon_0\ ,\nn\\
{\rm or} &&
\epsilon=e^{\ft{1}{2}\, r\, \hat\gamma\, \hat \Gamma_r}\, \Big( 1 +
\ft12 x(
\hat\gamma\, \hat\Gamma_x + \hat\Gamma_r\,\hat\Gamma_x)\Big)\,
e^{\ft{1}2\, \theta_2\, \hat\gamma\, \hat\Gamma_2}\,
e^{-\ft12\theta_1\, \hat\Gamma_{12}}\,
\epsilon_0\ ,
\eea
where the Dirac matrices are the ones appropriate to the total
spacetime dimension in each case.  In the case where one or other
space in the factored product is even dimensional, $\hat\gamma$ is the
pseudo chirality operator given by the appropriate product of the
hatted Dirac matrices in the even-dimensional factor.  For this
reason, there are two possibilities in the AdS$_2\times S^2$ example,
reflecting the two possibilities for the Dirac matrix decomposition
given in appendix A.  The first corresponds to taking $\hat\gamma$ to
be the pseudo chirality operator in AdS$_2$, and the second to taking
it instead to be in $S^2$.  In the case of AdS$_3\times S^3$, $\hat
\gamma \equiv \hat\Gamma^{123}=-\im\,
\sigma_2\otimes\oneone\otimes\oneone$, where the numerical indices lie
in $S^3$, while $\hat{\td\gamma} \equiv\ft16 \epsilon_{\mu\nu\rho}\,
\hat\Gamma^{\mu\nu\rho}=\sigma_1\otimes\oneone\otimes\oneone$.  In all
the examples, $\epsilon_0$ is an arbitrary constant spinor in the
total space.  It will be subject to a chirality condition in the
AdS$_3\times S^3$ example, if the $D=6$ supergravity is chosen to be
the minimal chiral theory, and $\hat{\td\gamma}$ can then be replaced
by $\hat\gamma$ in the expression for the Killing spinors.

\section{Discussion}

    In this paper, we have obtained explicit expressions for the
Killing spinors on $S^n$ for all $n$.  We then used the results to
obtain the full Killing spinors on various AdS$_m\times S^n$
spacetimes that arise as solutions in supergravity theories.  These
are of considerable interest owing to the conjectured duality relation
to conformal theories on the AdS boundaries.  One further application
of these results is to construct the Killing vectors, and conformal
Killing vectors, from appropriate bilinear products
${\epsilon'}^\dagger\, \Gamma_i\, \epsilon$ of Killing spinors.  As
discussed in \cite{fy}, products where the Killing spinors $\epsilon'$
and $\epsilon$ on $S^n$ either both satisfy (\ref{ks1}) or both
satisfy (\ref{ks1m}) give Killing vectors, while products where one
satisfies (\ref{ks1}) and the other satisfies (\ref{ks1m}) give
conformal Killing vectors.  In general, it is necessary to use both of
the Killing-vector constructions in order to obtain all the Killing
vectors on $S^n$.  At large $n$ there is a considerable redundancy in
the construction, since the number of Killing spinors grows
exponentially with $n$, while the number of Killing vectors grows only
quadratically with $n$.  In certain low dimensions, there is a more
elegant exact spanning of the Killing vectors using this construction,
such as for $S^7$ where the antisymmetric products $\bar\epsilon^\a\,
\Gamma_i\, \epsilon^\beta$ of the eight Killing spinors $\epsilon^\a$
give the 28 Killing vectors of $SO(8)$ \cite{dnp}.

    We shall present just one simple example here, for the case of
$S^2$.  From the matrix expression (\ref{om2}) in the appendix, we
find that from the Killing spinors $\epsilon= \Omega_2\, \pmatrix{a
\cr b}$ and $\epsilon'= \Omega_2\, \pmatrix{a' \cr b'}$, we obtain the
Killing vectors
\bea
K &=& K^i\, \del_i= E^i_j\, {\epsilon'}^\dagger\, \Gamma^j\, \epsilon
\nn\\
&=& (b \bar b' -a \bar a') \fft{\del}{\del\theta_1} +
\im\, (a \bar b' e^{-\im\theta_1} - \bar a' b e^{\im\theta_1})\,
\fft{\del}{\del\theta_2} +
(a \bar b' e^{-\im\theta_1} +\bar a' b e^{\im\theta_1})\,
\cot\theta_2\, \fft{\del}{\del\theta_1}\ ,
\eea
where $E^i_j$ are the components of the inverse vielbein $E_j=E^i_j\,
\del_i$.
Choosing different values for the constants $a, b,a',b'$ spans the
complete set of three Killing vectors of $SO(3)$.


\section*{Acknowledgment}

   We are grateful to Renata Kallosh for posing the question of
whether simple explicit expressions for the Killing spinors on $S^n$
can be obtained.  J.R. thanks Arvind Rajaraman for useful discussions.

\appendix

\section{Dirac matrices and their decomposition on product spaces}

    It is useful in general to represent the Dirac matrices in terms
of the $2\times 2$ Pauli matrices $\{\sigma_1, \sigma_2,\sigma_3\}$
as follows.  In even dimensions $D=2n$, we have
\bea
\Gamma_1 &=& \sigma_1 \otimes \oneone \otimes \oneone \otimes \cdots
\otimes \oneone\ ,\nn\\
\Gamma_2 &=& \sigma_2 \otimes \oneone \otimes \oneone \otimes \cdots
\otimes \oneone\ ,\nn\\
\Gamma_3 &=& \sigma_3 \otimes \sigma_1 \otimes \oneone \otimes \cdots
\otimes \oneone\ ,\nn\\
\Gamma_4 &=& \sigma_3 \otimes \sigma_2 \otimes \oneone \otimes \cdots
\otimes \oneone\ ,\nn\\
\Gamma_5 &=& \sigma_3 \otimes \sigma_3 \otimes \sigma_1 \otimes \cdots
\otimes \oneone\ ,\nn\\
&&\cdots \cdots \ ,\nn\\
\Gamma_{2n-1} &=& \sigma_3\otimes\sigma_3\otimes\cdots
\otimes\sigma_3\otimes \sigma_1\ ,\nn\\
\Gamma_{2n} &=& \sigma_3\otimes\sigma_3\otimes\cdots
\otimes\sigma_3\otimes \sigma_2\ ,\label{gammaeven}
\eea
In odd dimensions $D=2n+1$, we use the above construction for the
Dirac matrices of $2n$ dimensions, and take
\be
\Gamma_{2n+1} = \sigma_3\otimes\sigma_3 \otimes \sigma_3\otimes \cdots
\otimes\sigma_3\ .\label{gammaodd}
\ee

   When performing Kaluza-Klein reductions, it is necessary to
decompose the Dirac matrices of $D$ dimensions in terms of those of
the lower-dimensional spacetime $M_m$, and the internal space $K_n$,
whose respective dimensions $m$ and $n$ add up to $D$.  There are
four cases that arise, namely $(m,n) =$ (even,odd), (odd,even),
(even,even) and (odd,odd).  If we denote the Dirac matrices of the
spacetime $M_m$ by $\Gamma_\mu$, and those of the internal space $K_n$
by $\Gamma_i$, then the Dirac matrices $\hat\Gamma_A$ of $M_m\times
K_n$ can be written as:
\bea
\hbox{(even,odd)}:&& \hat\Gamma_\mu = \Gamma_\mu \otimes \oneone\ ,
 \qquad \hat \Gamma_i = \gamma\otimes\Gamma_i\ ,\nn\\
\hbox{(odd,even)}:&& \hat\Gamma_\mu = \Gamma_\mu \otimes \gamma\ ,
 \qquad \hat \Gamma_i = \oneone\otimes\Gamma_i\ ,\nn\\
\hbox{(even,even)}:&& \hat\Gamma_\mu = \Gamma_\mu \otimes \oneone\ ,
 \qquad \hat \Gamma_i = \gamma\otimes\Gamma_i\ ,\nn\\
\hbox{or} &&\hat\Gamma_\mu = \Gamma_\mu \otimes \gamma\ ,
 \qquad \hat \Gamma_i = \oneone\otimes\Gamma_i\ ,\nn\\
\hbox{(odd,odd)}:&& \hat\Gamma_\mu = \sigma_1\otimes\Gamma_\mu
\otimes \oneone\ ,
 \qquad \hat \Gamma_i = \sigma_2\otimes \oneone\otimes\Gamma_i\ ,
\eea
Note that in the final case the extra Pauli matrices $\sigma_1$ and
$\sigma_2$ are needed in order to satisfy the Clifford algebra, in
view of the fact that the Dirac matrices
of $D$ dimensions are twice the size of the simple tensor products of
those in $M_m$ and $K_n$.  Note also in this case that the chirality
operator in the total space is $\sigma_3\otimes\oneone\times\oneone$.

\section{Some low-dimensional examples}

    In this appendix, we give explicit matrix expressions for the
Killing spinors on the spheres $S^2$, $S^3$, $S^4$ and $S^5$.  These
examples arise in the near-horizon structures of
Rei{\ss}ner-Nordstr{\o}m
black holes, dyonic strings, M5-branes and D3-branes respectively.  In
each case, we may write the expression (\ref{sol1}) for the Killing
spinors on $S^n$ as $\epsilon =\Omega_n\, \epsilon_0$.  For $S^2$,
taking $\Gamma_i=\sigma_i$, where $\sigma_i$ are the usual Pauli
matrices, we find
\be
\Omega_2 = \pmatrix{e^{-\ft{\im}{2}\theta_1}\, \cos\ft12\theta_2 &
                   e^{\ft{\im}{2}\theta_1}\, \sin\ft12\theta_2\cr
                 -e^{-\ft{\im}{2}\theta_1}\, \sin\ft12\theta_2 &
               e^{\ft{\im}{2}\theta_1}\, \cos\ft12\theta_2}
\ .\label{om2}
\ee
To avoid clumsy expressions later, we may define
$t_k =e^{\ft{\im}{2}\theta_k}$, $\bar t_k =e^{-\ft{\im}{2}\theta_k}$,
$c_k=\cos\ft12\theta_k$, $s_k=\sin\ft12\theta_k$.  The matrix
$\Omega_2$ thus becomes
\be
\Omega_2 =\pmatrix{ \bar t_1\, c_2 & t_1\, s_2 \cr
                    -\bar t_1\, s_2 & t_1\, c_2}\ .
\ee
For $S^3$, $S^4$ and $S^5$ we obtain
\bea
\Omega_3 &=&\pmatrix{ \bar t_1\, t_3\, c_2 & -\im
\, t_1\, t_3\, s_2 \cr
        -\im\, \bar t_1\, \bar t_3\, s_2 & t_1\,
\bar t_3 \, c_2}\ ,\\
&&\nn\\
\Omega_4 &=& \pmatrix{
\bar t_1 \,\bar t_3\, c_2\, c_4
& \bar t_1 \,  \bar t_3\, c_2\, s_4&
-\im\, t_1\, \bar t_3\, s_2\, s_4 &
-\im\, t_1\, \bar t_3\, c_4\, s_2\cr
- \bar t_1 \,  \bar t_3\, c_2\, s_4
&  \bar t_1 \,  \bar t_3\, c_2\, c_4
&-\im\, t_1\, \bar t_3 \, c_4\, s_2
& \im\, t_1\, \bar t_3\, s_2\, s_4\cr
\im\,  \bar t_1 \,  t_3\, s_2\, s_4
&-\im\,  \bar t_1 \,  t_3\, c_4\, s_2
& t_1 \,  t_3\, c_2\, c_4
&-  t_1 \,  t_3\, c_2\, s_4\cr
-\im\,  \bar t_1 \,  t_3\, c_4\, s_2
& -\im\,  \bar t_1 \,  t_3\, s_2\, s_4
&t_1 \,  t_3\, c_2\, s_4
&  t_1 \,  t_3\, c_2\, c_4}\ ,\\
&&\nn\\
\Omega_5 &=&
\pmatrix{
\bar t_1 \,  \bar t_3 \,  \bar t_5\, c_2\, c_4
& -\im\,  \bar t_1 \,  \bar t_3 \,  \bar t_5\, c_2\, s_4
&- t_1\, \bar t_3 \,  \bar t_5 \, s_2\, s_4
& -\im\,  t_1\, \bar t_3 \,  \bar t_5 \, c_4\, s_2\cr
-\im\,  \bar t_1 \,  \bar t_3 \,  \bar t_5\, c_2\, s_4
& \bar t_1 \,  \bar t_3 \,  \bar t_5\, c_2\, c_4
& -\im\, t_1\,  \bar t_3 \,  \bar t_5 \, c_4\, s_2
& -t_1\, \bar t_3 \,  \bar t_5 \,  s_2\, s_4\cr
-\bar t_1 \,  t_3 \,  t_5\, s_2\, s_4
& -\im\,  \bar t_1 \,  t_3 \,  t_5\, c_4\, s_2
& t_1 \,  t_3 \,  t_5\, c_2\, c_4
& -\im\,  t_1 \,  t_3 \,  t_5\, c_2\, s_4\cr
-\im\,  \bar t_1 \,  t_3 \,  t_5\, c_4\, s_2
& -  \bar t_1 \,  t_3 \,  t_5\, s_2\, s_4
& -\im\,  t_1 \,  t_3 \,  t_5\, c_2\, s_4
&  t_1 \,  t_3 \,  t_5\, c_2\, c_4}\ .
\eea

    In these examples we have used the representations of Dirac
matrices given in equations (\ref{gammaeven}) and (\ref{gammaodd})
of appendix A.

\end{document}